\documentclass[a4paper,12pt]{article}
 \pdfoutput=1
\usepackage{graphicx,rotating,hyperref,slashed,amsmath,xcolor,catchfilebetweentags,ifluatex,cite,fourier,multirow,expdlist}

\makeatletter
\hypersetup{colorlinks,bookmarksopen,bookmarksnumbered,
linkcolor=blus,pdfstartview=FitH,urlcolor=rossos,citecolor=verde}

\newcommand{\np}{^{(n)}} 
\newcommand{\kpn}{\ket{\psi\np}}

 \renewcommand{\Im}{{\rm Im}\,}
\renewcommand{\Re}{{\rm Re}\,}
\newcommand{\mio}[1]{}
\newcommand{\new}[1]{{\color{blue}#1}}
\renewcommand{\new}[1]{#1}
\newcommand{\xxx}[1]{{\color{red}[\bf #1]}}
\newcommand{\yyy}[1]{{\color{blue}\sl#1}}
\newcommand{\fig}[1]{~\ref{fig:#1}}

\newcommand{\bra}[1]{\langle #1 |}
\newcommand{\ket}[1]{| #1 \rangle}
\newcommand{\bk}[2]{\langle #1  |  #2  \rangle}
\newcommand{\kb}[2]{| #1  \rangle\!\langle   #2  |}
\newcommand{\bAk}[3]{\langle #1  |  #2|#3  \rangle}
\definecolor{Gray}{gray}{0.95}

\newcommand{\sfrac}[2]{#1/#2}

\usepackage{multicol}
\usepackage{color}
\definecolor{rosso}{cmyk}{0,1,1,0.4}
\definecolor{rossos}{cmyk}{0,1,1,0.55}
\definecolor{rossoc}{cmyk}{0,1,1,0.2}
\definecolor{blu}{cmyk}{1,1,0,0.3}
\definecolor{blus}{cmyk}{1,1,0,0.6}
\definecolor{bluc}{cmyk}{1,1,0,0.1}
\definecolor{verde}{cmyk}{0.92,0,0.59,0.25}
\definecolor{verdec}{cmyk}{0.92,0,0.59,0.15}
\definecolor{verdes}{cmyk}{0.92,0,0.59,0.4}

\newcommand{\eq}[1]{~{\rm (\ref{eq:#1})}}

\newcommand{\GeV}{\,{\rm GeV}}

\newcommand{\diag}{\,{\rm diag}}

\def\circa#1{\,\raise.3ex\hbox{$#1$\kern-.75em\lower1ex\hbox{$\sim$}}\,}

\newcommand{\beq}{\begin{equation}}
\newcommand{\eeq}{\end{equation}}

\newcommand{\bea}{\begin{eqnarray}}
\newcommand{\eea}{\end{eqnarray}}
\newcommand{\be}{\begin{equation}}
\newcommand{\ee}{\end{equation}}
\font\tenrsfs=rsfs10 at 12pt
\font\sevenrsfs=rsfs7 at 10 pt
\font\fiversfs=rsfs5
\newfam\rsfsfam
\textfont\rsfsfam=\tenrsfs
\scriptfont\rsfsfam=\sevenrsfs
\scriptscriptfont\rsfsfam=\fiversfs
\def\mathscr#1{{\fam\rsfsfam\relax#1}}

\def\circa#1{\,\raise.3ex\hbox{$#1$\kern-.75em\lower1ex\hbox{$\sim$}}\,}
\makeatletter

\def\hhref#1{\href{http://arxiv.org/abs/#1}{arXiv:#1}} 

\newcommand{\ital}{\em}
\def\hhref#1{\href{http://arxiv.org/abs/#1}{arXiv:#1}}
\usepackage{xstring} 
\newcommand{\hhrefq}[1]{\IfSubStr{#1}{:}{\href{http://inspirehep.net/search?ln=en&ln=en&p=#1&of=hb&action_search=Search&sf=&so=d&rm=&rg=25&sc=0}{InSpire:#1}}{\hhref{#1}}}

\def\art{\@ifnextchar[{\eart}{\oart}}
\def\eart[#1]#2#3#4#5#6{{\rm #2}, {\em #3 \bf #4} {\rm (#6) #5} ({\em #1})}
\def\article{\@ifnextchar[{\earticle}{\oarticle}}
\def\oarticle#1#2#3#4#5#6{{\rm #1}, {\ital ``#6''}, {\rm #2 #3 (#5) #4}}
\def\earticle[#1]#2#3#4#5#6#7{{\rm #2}, {\ital ``#7''}, {\rm #3 #4 (#6) #5}  [\hhrefq{#1}]}
\def\hepart[#1]#2{{\rm #2, \sl#1}}
\def\heparticle[#1]#2#3{#2, {\ital ``#3''} [\hhrefq{#1}]}
\newcommand{\doi}[1]{\href{http://dx.doi.org/#1}{[link]}}

\newcounter{alphaequation}[equation]
\def\thealphaequation{\theequation\hbox to
0.6em{\hfil\alph{alphaequation}\hfil}}
\def\eqnsystem#1{
\def\@eqnnum{{\rm (\thealphaequation)}}
\def\@@eqncr{\let\@tempa\relax \ifcase\@eqcnt \def\@tempa{& & &} \or
  \def\@tempa{& &}\or \def\@tempa{&}\fi\@tempa
  \if@eqnsw\@eqnnum\refstepcounter{alphaequation}\fi
\global\@eqnswtrue\global\@eqcnt=0\cr}
\refstepcounter{equation} \let\@currentlabel\theequation \def\@tempb{#1}
\ifx\@tempb\empty\else\label{#1}\fi
\refstepcounter{alphaequation}
\let\@currentlabel\thealphaequation
\global\@eqnswtrue\global\@eqcnt=0 \tabskip\@centering\let\\=\@eqncr
$$\halign to \displaywidth\bgroup \@eqnsel\hskip\@centering
$\displaystyle\tabskip\z@{##}$&\global\@eqcnt\@ne
\hskip2\arraycolsep\hfil${##}$\hfil& \global\@eqcnt\tw@\hskip2\arraycolsep
$\displaystyle\tabskip\z@{##}$\hfil
\tabskip\@centering&\llap{##}\tabskip\z@\cr}
\def\endeqnsystem{\@@eqncr\egroup$$\global\@ignoretrue}

      \makeatother
      
\begin{document}


\vspace{1cm}

\begin{center}
{\LARGE \bf \color{rossos}
Interpretation of quantum  \\[2mm]
mechanics with indefinite norm
}\\[1cm]

{\large\bf Alessandro Strumia}  
\\[7mm]
{\it CERN, Theory Division, Geneva, Switzerland\\
Dipartimento di Fisica dell'Universit{\`a} di Pisa and INFN, Italia\\
}

\vspace{1cm}
{\large\bf\color{blus} Abstract}
\begin{quote}
The Born postulate can be reduced to its deterministic content that only applies to eigenvectors of observables:
the standard probabilistic interpretation of generic states then follows from algebraic properties of repeated measurements and states.
Extending this reasoning suggests an interpretation of quantum mechanics generalized with  indefinite quantum norm.
\end{quote}

\thispagestyle{empty}
\end{center}
\begin{quote}
{\large\noindent\color{blus} 
}

\end{quote}
\tableofcontents

\setcounter{footnote}{0}


\section{Introduction}
If a gravitational action with 4 derivatives leads to 
a sensible quantum theory,
the resulting quantum gravity
has welcome properties:  renormalizability~\cite{Stelle:1976gc,1403.4226},
inflation for generic dimension-less potentials~\cite{1403.4226}, dynamical generation of a naturally small electro-weak scale~\cite{1403.4226,1705.03896}.
Quantum fields can be expanded in modes, motivating the 
study of the basic building block: one variable $q(t)$ with 4 derivatives.
In the canonical formalism, it can be rewritten in terms of two variables with 2 derivatives, $q_1 = q$ and $q_2 = d q/dt$.
The classical theory \new{tends to have run-away solutions}, 
because the classical Hamiltonian is unbounded from below~\cite{Ostro}
\new{(although instabilities do not take place  for some range of initial conditions and/or in special systems~\cite{AntiOstro}).}
However nature is quantum. Systems with 1 derivative (fermions) provide an example
where the same problem --- a classical Hamiltonian unbounded from below --- is not present in the quantum theory.

This motivates the study of quantisation of systems with 4 derivatives.
In view of the time derivative, $q_1$ and $q_2$ have opposite time-inversion parities.
This is satisfied using for $q_{1,2}$ the two different coordinate representations of
a pair of canonical coordinates with $[\hat q,\hat p]=i$:
$$\begin{array}{c|cc|cc|c}
\hbox{proposed by} & \bAk{x}{\hat q}{\psi} & T\hbox{-parity} &  \bAk{x}{\hat p}{\psi} & T\hbox{-parity} & \hbox{norm $\bk{x'}{x}$}\\ \hline
\hbox{Schroedinger}  &x\psi(x) & \hbox{even} & -i\,d\psi/d x & \hbox{odd} &\hbox{$\delta(x-x')$, positive}\\
\hbox{Dirac-Pauli}  &-ix\psi(x) & \hbox{odd} & d\psi/d x & \hbox{even} &\hbox{$\delta(x+x')$, indefinite}
\end{array}$$
The first possibility is the well known positive-norm Schroedinger representation.
The second possibility, first described by Dirac~\cite{Dirac} and studied by Pauli~\cite{Pauli}, 
remained less known because  $\hat q$ and $\hat p$ are self-adjoint under
the indefinite quantum norm $\bk{x'}{x} = \delta(x+x')$.\footnote{In mathematical convention
the norm is, by definition, positive and one should speak of ``inner product'' in ``Krein space''.
We avoid using these terms, keeping the standard terms of quantum mechanics.}
The modified time-reflection $T$-parity comes from the unusual $i$ factor.

The 4-derivative oscillator $q(t)$, quantised 
proceeding along these lines as  $\hat {q}_1\ket{q_1,q_2} = q_1 \ket{q_1, q_2}$
and $\hat{q}_2 \ket{q_1,q_2}  = i q_2 \ket{q_1,q_2} $,
leads to a successful formalism similar to the deterministic part of  quantum mechanics:
positive energy eigenvalues, normalizable wave-functions,
a time evolution $\hat U =e^{-i\hat Ht}$ which conserves the indefinite quantum norm if the Hamiltonian $\hat H$ is
self-adjoint~\cite{1512.01237}.

The remaining problem is whether such formalism admits a physical interpretation.
In the conventional interpretation of quantum mechanics, positive norm 
is interpreted as probability of outcomes of measurements.
In section~\ref{neversayprob} we describe a rephrasing of conventional quantum mechanics inspired by~\cite{Nima},
where probability is replaced by average over many repeated measurements.
Then the full probabilistic Born rule follows from its deterministic part,
combined with the algebraic properties of repeated quantum states.
In section~\ref{negnorm} we apply the deterministic part of the Born postulate
to  indefinite-norm quantum mechanics, finding the implied interpretation.
Examples are given in section~\ref{examples}. 
Results are summarized in the conclusions, given  in section~\ref{concl}.

Various authors explored possible interpretations of indefinite-norm quantum mechanics (or equivalently
of pseudo-hermitian hamiltonians~\cite{Bender})
trough algebraic approaches that construct one artificial arbitrary positive norm
choosing the special basis of Hamiltonian eigenstates~\cite{Bender,quant-ph/0609032,Mosta,1611.03498}.
Their definition is similar to our final result, except that 
in our approach each observable selects the basis of its own eigenstates.

\section{Quantum mechanics bypassing probabilities}\label{neversayprob}
The Born postulate 
\begin{quote}\em
``when an observable corresponding to a self-adjoint operator $\hat{A}$ is measured in a state $\ket{\psi}$,
the result is an eigenvalue $A_i$ of $\hat{A}$ with probability 
$ p_i= |\bk{A_i}{\psi}|^2$
\end{quote}
employs probability only if $\ket{\psi}$ is a generic state.
If instead $\ket{\psi}$ is  an eigenstate of the operator to be measured the probability is unity, which means certainty:
the Born rule reduces to the following deterministic statement:
\begin{quote}\em
``when an observable corresponding to a self-adjoint operator $\hat{A}$ is measured in an eigenstate $\ket{A_i}$ of $\hat{A}$,
the result is the eigenvalue $A_i$''.
\end{quote}
As discussed below, this is enough to make useful predictions even for non-trivial states.
The point is that a probability $p$ about one measurement
can be rephrased as a certainty about repeated experiments.
This is what quantum experimentalists do: repeat a measurement 
$n\gg 1$ times; the outcome of the measurements is their average.
The repeated measurements can be done at different times (for example when measuring a cross section at a collider) or at different places
(for example when observing a primordial cosmological inhomogeneity `measured' by the early universe).
A useful formalism that avoids such details and describes  a single measurement repeated $n$ times
consists in defining  a state $\kpn  $ equal to the tensor product $\ket{\psi}\otimes \cdots \otimes \ket{\psi}$
of $n$ identical copies of the generic state $\ket{\psi}$
subject to the measurement.
We will see  that  $\kpn  $ becomes,
in the limit
$n\to \infty$, an eigenstate of the operator
$\hat{A}\np$ (constructed later) that describes  a  measurement repeated $n$ times.
Similar ideas have been presented in~\cite{Nima}.

\subsection{Repeated states}

It is convenient to use a basis of eigenstates $\ket{A_i}$ of the observable $\hat{A}$ normalized as $\bk{A_i}{A_j}=\delta_{ij}$.
To start, a state $\ket{\psi} = c_1 \ket{A_1} + c_2 \ket{A_2}$ repeated twice becomes
\beq |\psi\rangle^{(2)} = c_1^2  \ket{1^22^0} 
+ c_2^2\ket{1^02^2}  + \sqrt{2}c_1 c_2 \ket{1^12^1}\eeq
where each coefficient multiplies a basis vector with unit norm in the tensor space:
\beq \ket{A_1^2  A_2^0} = \ket{A_1}\ket{A_1},\qquad
\ket{A_1^0  A_2^2}=\ket{A_2}\ket{A_2},\qquad
\ket{A_1^1 A_2^1}=\frac{\ket{A_1} \ket{A_2} + \ket{A_2} \ket{A_1}}{\sqrt{2}}.\eeq
The generalization of the factor $\sqrt{2}$ in the third term is important in the following.
Higher powers $ \kpn  $ of a generic state $\ket{\psi} = \sum_{i=1}^N c_i \ket{A_i}$ can be written as
\beq \label{eq:psin}\kpn   = \sum_{k_1 +\cdots + k_N = n} c_1^{k_1} \cdots c_N^{k_N} \sqrt{{n\choose k_1\cdots k_N}}
\ket{A_1^{k_1}\cdots A_N^{k_N}}\eeq
where 
\beq {n \choose k_1\cdots k_N}   = \frac{n!}{k_1!\cdots k_N!}\eeq
is the  multinomial coefficient. The basis states are
\beq  
\ket{A_1^{k_1} \cdots A_N^{k_N}} =  \frac{\sum_{\rm perm} |A_1\rangle^{k_1} \cdots \ket{A_N}^{k_N} }{\sqrt{{n\choose k_1\cdots k_N}}}
\eeq
where the sum runs over all permutations.  Such states are normalized as
\beq
\bk{A_1^{k_1}\cdots A_N^{k_N}}{A_1^{k_1} \cdots A_N^{k_N}} = \bk{A_1}{A_1}^{k_1} \cdots \bk{A_N}{A_N}^{k_N}
=1.\eeq
Thereby
$ {}\bk{\psi\np}{\psi\np}  = \bk{\psi}{\psi}^n $
equals unity provided that $ \bk{\psi}{\psi} = \sum_{j=1}^N  |c_j^2| = 1$.

\medskip

\begin{figure}[t]
$$\includegraphics[width=0.7\textwidth]{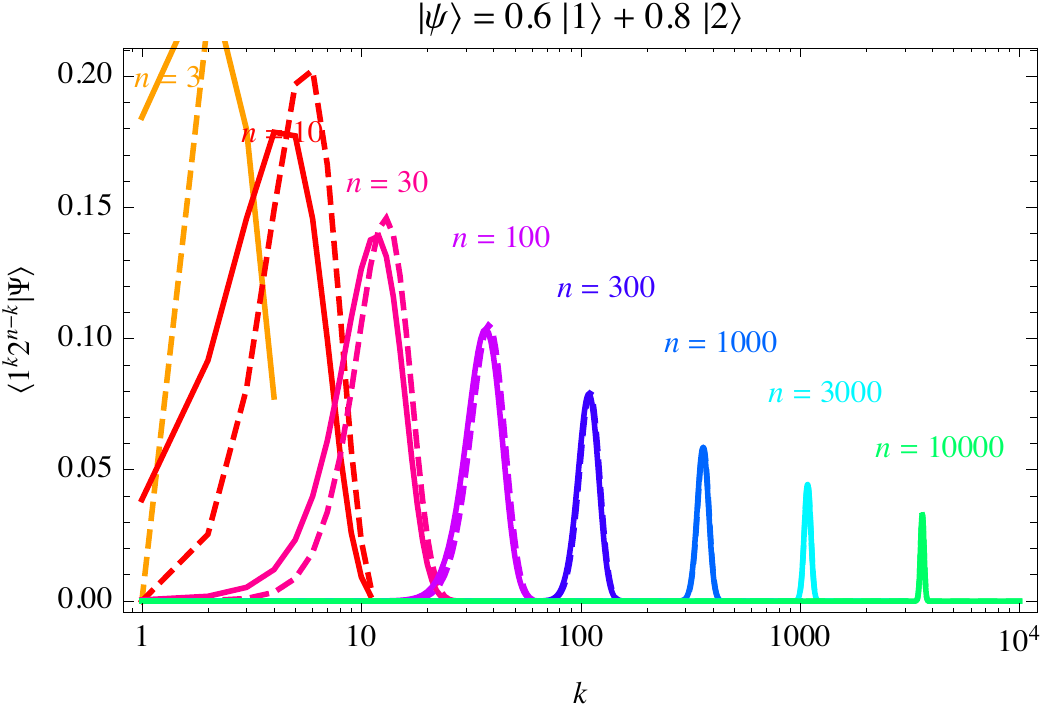}$$
\caption{\label{fig:bells}\em We examplify how a pure state is obtained from high powers $n$ of
a generic state $\ket{\psi}= c_1\ket{1} + c_2 \ket{2}$.
The continuous bells are the coefficients of the state $\ket{\Psi}=p_1\kpn  $ along the basis states $\ket{1^k2^{n-k}}$ for different values of the power $n$.
Furthermore, the dashed bells are  the coefficients of
$\ket{\Psi}=\hat P_1 \kpn  $, where $\hat P_1$ is the observable that counts the rate of $\ket{1}$.
We see that
for large $n$ both distributions approach a common narrow bell,
meaning that $\kpn  $ is an eigenvalue of $\hat P_1$ with eigenvalue $p_1 = |c_1^2|$.
This is the standard result of quantum mechanics, obtained without using probabilities.
}
\end{figure}

It is convenient to split the coefficients as $c_i = e^{i\delta_i} |c_i|$ and rewrite eq.\eq{psin} as
\beq\label{eq:sqrtmulti}
 \kpn   = \sum_{k_1 +\cdots + k_N = n} e^{i (k_1 \delta_1 + \cdots + k_N \delta_N)}
\sqrt{|c_1^2|^{k_1} \cdots |c_N^2|^{k_N}  {n\choose k_1\cdots k_N}}\,
\ket{A_1^{k_1}\cdots A_N^{k_N}}.\eeq
All the phases could be set to $\delta_i =0$ trough a re-phasing of the eigenstates $\ket{i}$.
The multinomial plays a double role in mathematics:
it is a tool for computing powers and it is also used in the multinomial probability distribution.  
The term under the square root has the same form as the multinomial distribution
for obtaining $k_i$ events of type $i$ in $n$ trials
with `probability coefficients'  \beq \label{eq:pi}
p_i = \frac{|c_i|^2}{\sum_j |c_j^2|}.\eeq
We never used probabilities: the multinomial distribution appeared by computing tensor powers
$\kpn  $, which manage
to result into coefficients $p_i$ proportional to moduli squared.\footnote{Ignoring the quantum state algebra and expanding $(c_1 + \cdots + c_N)^n $ would have given instead $p_i = c_i/(\sum_j c_j)$.}

As well known, in the limit $n\gg 1$ one can approximate
$n!\simeq n^n  e^{-n} \sqrt{2\pi n}[1+{\cal O}(1/n)]$
finding that a multinomial reduces to a Gaussian
with mean $\mu_i  = n p_i$ and variance
$\sigma_{ij}^2 = n (p_i \delta_{ij} - p_i p_j)$.
In the limit $n\to \infty$ the standard deviation becomes negligible with respect to the mean,
and the Gaussian further reduces to a Dirac delta, $\delta(k_i - \mu_i)$.
Fig.~\ref{fig:bells} shows an example for a 2-state system.

\subsection{Repeated measurements}
The above discussion suggests that for $n\to\infty$ the state $\kpn$ 
can become an eigenstate of appropriate observables that do not probe the detailed structure
of the narrow peak, which contains $\sim \sqrt{n}$ states. 
Then the deterministic part of the Born rule predicts the outcome of the measurement.
The appropriate observable is the average of a generic single-state observable $\hat{A}$ over $n$ repeated measurements.
Formally, such repeated measurements are described by operators $\hat{A}\np$ of the form 
\beq\label{eq:An}
\hat{A}^{(1)}=\hat{A},\qquad
\hat{A}^{(2)} = \frac{\hat{A}\otimes \hat 1 + \hat 1\otimes \hat{A}}{2} ,\qquad
\hat{A}^{(3)} = \frac{\hat{A}\otimes \hat 1 \otimes \hat 1 +\hat  1\otimes \hat{A} \otimes \hat 1 +\hat  1\otimes \hat 1 \otimes \hat{A}}{3}, \eeq
etc.  Indeed their averages satisfy
\beq \label{eq:Anavg}
 \frac{\bAk{\psi\np}{\hat{A}\np}{\psi\np}  }{\bk{\psi\np}{\psi\np}}=  \frac{\bAk{\psi}{\hat{A}}{\psi}}{\bk{\psi}{\psi}}.\eeq
One basic observable is the projector $\hat \Pi_i$ over the state $\ket{A_i}$.
Acting over tensored space, $\hat P_i \equiv \hat \Pi_i^{(n)}$ counts the rate of $\ket{A_i}$ states:
\beq \label{eq:Pi}
\hat P_i \ket{A_1^{k_1}\cdots A_N^{k_N}} = \frac{k_i}{n} \ket{A_1^{k_1}\cdots A_N^{k_N}}.\eeq
This is just a way of formalizing what experimentalists do.
For example, when measuring the component of the spin of a fermion in the state
$\ket{\psi} = c_\uparrow \ket{\uparrow} + c_\downarrow \ket{\downarrow}$,
the experimentalist 
builds an  apparatus with magnetic fields that deflect quanta trough a spin-dependent force,
such that the whole system (status plus apparatus) evolves into
$c_\uparrow \ket{\uparrow} \ket{{\rm up-going}}+ c_\downarrow \ket{\downarrow} \ket{\rm down-going}$,
where $\ket{\rm up-going}$ and $\ket{\rm down-going}$ are the macroscopic
states with a stable entanglement with the apparatus.
Being macroscopic, their relative phase oscillates fast averaging to zero:
the apparatus forced the status to `choose'  among its components.
Any single observation gives either up or down forming a random sequence
from which the experimentalist can extract the rate $p_\uparrow$ of ups.

\smallskip

Fig.~\ref{fig:bells} exemplifies how $p_i \kpn  $ and $\hat P_i \kpn  $ converge to the same state for large $n$.
We consider a two-state system with norm $\bk{\psi}{\psi}=1$,
such that the basis states of $\kpn  $ are $\ket{1^k 2^{n-k}}$, described by one integer $k$ than runs from 0 to $n$.
The eigenvalue $p_i$ is determined to be as in eq.\eq{pi}
by imposing that the coefficients of the two states 
\beq c_k^{(n,i)} = \bAk{1^k 2^{n-k}}{p_i}{\psi\np} ,\qquad
C^{(n,i)}_k= \bAk{1^k 2^{n-k}}{\hat P_i}{\psi\np} \eeq
are equal on the peak at $k \approx n p_i$.
For large $n$ the peak gets so relatively narrow 
(the average grows as $n$, the width as $\sqrt{n}$)
that all other values of $k$ are irrelevant.
The situation is intuitively clear for large finite $n$,
and nobody repeats experiments an infinite number of times.
Nevertheless it is interesting to discuss the  limit $n\to\infty$,
showing two possible notions of convergence:
\begin{itemize}

\item[C)] {\bf Coefficient convergence}.
Fig.~\ref{fig:bells} shows that the height of the peak decreases with $n$, so
both sequence of coefficients $c_k^{(n,i)}$ and $C^{(n,i)}_k$ tend to zero as $n\to\infty$.
This would be exacerbated by choosing $\bk{\psi}{\psi}<1$, while
$\bk{\psi}{\psi}>1$ would make all coefficients divergent as $n\to\infty$.
Given that we want to compute eigenvectors and eigenvalues, 
the normalization of states is irrelevant 
(as usual in quantum mechanics), so the right notion of
convergence is projective:
\beq \label{eq:proCc}\lim_{n\to \infty}  \frac{c^{(n,i)}_k -C^{ (n,i)}_k}{\max_{k'} C^{(n,i)}_{k'}}= 0\eeq
which is satisfied.
In general this applies to a set of $k_j$, rather than a single $k$.

\item[N)]  {\bf Norm convergence}.
The norm of
$(\hat P_i - p_i)\kpn$ converges projectively to zero
\beq\lim_{n\to\infty} \bAk{\psi\np}{ (\hat P_i - p_i)^2}{ \psi\np} /\bk{\psi\np}{\psi\np}=0\eeq
as can be proofed making use of~\cite{Nima} 
\beq \label{eq:pi+}
\frac{ \bAk{\psi\np}{\hat P_i}{\psi\np}  }{  {}\bk{\psi\np}{\psi\np}  }= p_i ,\qquad
 \lim_{n\to \infty} \frac{  \bAk{\psi\np}{\hat P_i^m}{\psi\np}  }{ {}\bk{\psi\np}{\psi\np}}  =p_i^m.\eeq 
\end{itemize}
Both notions of convergence lead to the same conclusion:
$ \kpn  $ becomes asymptotically an eigenstate of the observable $\hat P_i$ with eigenvalue $p_i$.
Then, the deterministic part of the Born rule predicts the outcome of the measurement to be the eigenvalue $p_i$.
This is the standard result of quantum mechanics, obtained replacing 
the probabilistic Born postulate with the average over many repeated measurements,
which is deterministically predicted.

For a generic observable $\hat{A}$ one asymptotically has $\hat{A}^{(n)}\kpn  =  \sum_i A_i p_i  \kpn $.
The commutator of two operators $\hat{A}$ and $\hat B$ satisfies $[\hat{A}^{(n)}, \hat B^{(n)}]=[\hat{A},\hat B]^{(n)}/n$, which gets suppressed
at large $n$, showing how quantum uncertainty reduces to classical determinism for $n\to\infty$.

\mio{By defining $\Delta A\np$ as $A\np$ minus its average value, 
and assuming $\bk{\psi}{\psi}=1$ for simplicity,
the corresponding uncertainty relation
\beq \sqrt{\bAk{\psi\np}{(\Delta A\np)^2}{\psi\np} \, \bAk{\psi\np}{(\Delta B\np)^2}{\psi\np}} \ge \frac{|\bAk{\psi\np}{[ A,B]\np}{\psi\np}|}{2n} = \frac{|\bAk{\psi}{[ A,B]}{\psi}|}{2n} 
\eeq
shows how classicality is obtained for large $n$.}

\medskip

One can next  introduce the concept of probability, giving a meaning to generic states $\ket{\psi}$.
But this is not necessary:
one can equivalently tell that  a generic state $\ket{\psi}$ has no physical meaning
given that a single measurement over it has a random outcome.

The important positive fact is that from any state $\ket{\psi}$ one can form a
pure state $\kpn  $ which has a deterministic physical meaning on 
a large class of operators of the form $\hat{A}\np$.

\medskip

Not all operators have a good classical $n\to\infty$ limit on multi-states $\ket{\psi\np}$.
An example of an alternative operator with a bad limit is
$\hat{A}^n = \hat{A}\otimes \hat{A} \otimes \hat{A}\otimes\cdots$ which measures the product
(rather than the average) of each single observation.
Consider e.g.\ the case where $\hat{A}$ is a parity with eigenvalues $\pm 1$, such as
$J_z$ in a Stern-Gerlach experiment.
The product parity of the system can flip sign with any extra measurement,
and does not lead to a useful observable in the limit $n\to \infty$.
Within the repeated-states formalism, this happens because
the eigenvalues of ${\hat{A}^n}$ wildly vary inside the $\sqrt{n}$ states within the peak,
such that $\ket{\psi\np}$ is an eigenvector of $\hat{A}\np$ but not of $\hat{A}^n$.
Similar issues will arise in the following section.

\section{Interpreting indefinite-norm quantum mechanics}\label{negnorm}
The previous discussion about consistency of repeated mesurments restricts possible interpretations of
standard quantum mechanics with positive quantum norm $\bk{\psi}{\psi}$,
allowing to (re)derive the Born  postulate from its deterministic part limited to eigenvectors. 
We here explore if the algebraic properties of repeated states allow to derive an interpretation of
 quantum mechanics generalized allowing for an indefinite norm.
What is the meaning of a state
$\ket{\psi} = c_+ \ket{+} + c_- \ket{-} $ which is a superposition of a positive-norm state $\ket{+}$
with a negative-norm state $\ket{-}$?
To answer, we divide observables in three classes, discussed in section~\ref{Acom}, \ref{Anocom} and~\ref{deg}.


\subsection{Observables that commute with one ghost operator}\label{Acom}
We  consider an observable described by a 
self-adjoint operator $\hat{A}$ in Krein space, and assume that 
its eigenvectors $\ket{A_i}$ univocally define a complete basis in configuration space.
Physically, this corresponds a good apparatus that converts orthogonal quantum states into different macroscopic states.
Mathematically, this means that the eigenvalues $A_i$  non-degenerate and that each eigenvector
lies away from the null cone in configuration space, such that the scalar products are
$\bk{A_i}{A_j} = N_i \delta_{ij}$ with $N_i =+1$ or $-1$.
The eigenvalues are then real.

\smallskip

From a generic state $\ket\psi = \sum_{i=1}^N c_i \ket{A_i}$ we again form
the repeated state $\kpn$ which becomes  an eigenstate of the repeated measurement
$\hat{A}\np$ in the limit of an infinite number of measurements, $n\to \infty$.
However the two notions of limit presented in section~\ref{neversayprob} for a positive norm
no longer give the same answer, so that a creative judgement is needed.
\begin{itemize}
\item[N)]  {\bf Norm convergence}.
The indefinite-norm averages of projectors 
considered in eq.\eq{pi+} are now given by
\beq    w_i \equiv
\frac{ \bAk{\psi\np}{\hat P_i }{\psi\np}  }{ \bk{\psi\np}{\psi\np}} =\frac{ N_i |c_i^2|}{\sum_j N_j |c_j^2|} ,\qquad
\lim_{n\to \infty} {}  \frac{\bAk{\psi\np}{\hat P_i^m }{\psi\np} }{  \bk{\psi\np}{\psi\np}}  =  w_i^m.\eeq
Thereby $ (\hat P_i-   w_i)\kpn $ asymptotically has zero projective norm,
\beq  \lim_{n\to \infty}  \frac{\bAk{\psi\np}{  (\hat P_i-   w_i)^2}{\psi\np}   }{ \bk{\psi\np}{\psi\np}}= 0. \eeq
Decomposing $\hat{A} = \sum_i A_i \hat \Pi_i$ in terms of eigenvalues $A_i$ and projectors $\hat\Pi_i$, 
$\kpn$ norm-converges to
 an eigenvector of $\hat{A}\np$ with eigenvalue $\sum_i A_i w_i$.
\end{itemize}
However, indefinite-norm convergence  does not imply convergence:
non-vanishing states along the null cone (such as $\ket+ + \ket-$) have null norm.
Furthermore, the $w_i$ coefficients can be negative such that they cannot interpreted as probabilities.
Finally, the averages contain huge cancellations, like in the expansion of $1 = (3-2)^{n}$ for large $n$.

\begin{itemize}
\item[C)] {\bf Coefficient convergence}.
The discussion in section~\ref{neversayprob} 
about the algebraic properties of $\kpn$ remains unaltered
 (in particular eq.s~\eq{psin}, \eq{An}, \eq{Pi}), up to one new issue:
the basis coefficients of $\kpn$ can get big and diverge even when computing high powers $\kpn$ of a unit-norm state such as
$\sqrt{3}\ket{+} + \sqrt{2}\ket{-}$.
Since the overall normalization of states has no physical meaning,  we renormalise the
coefficients of $\kpn$, for example setting the biggest coefficient to unity (as already done in eq.\eq{proCc}
to deal with positive norm and $\bk{\psi}{\psi}>1$).
Following this intuitive procedure one finds that
for large $n$ the coefficients of $\kpn$
again projectively converge to a narrow bell peaked at the same $p_i$ as in eq.\eq{pi},
\beq p_i  = 
\frac{|c_i^2|}{\sum_j |c_j^2|},\qquad 0 \le p_i \le 1 \eeq
and the coefficients of $\hat P_i\kpn$ converge to those of $p_i\kpn$.
Thereby $\kpn$ becomes eigenvector of $\hat{A}\np$ with eigenvalue $\sum_i A_i p_i$.
\end{itemize}
Choosing coefficient convergence,
the deterministic part of the Born postulate implies that  
$\sum_i A_i p_i$  must be identified with the measured value of $\hat{A}\np$,
and again the $p_i$ can be considered as probabilities.

\medskip

The intuitive procedure of renormalising a state repeated $n$ times 
before taking the limit $n\to\infty$
can be put on a more solid formal basis by defining an artificial positive norm  as follows. 
We define a {\em `ghost operator'} $\hat G$ as any linear operator such that $\hat G \ket i = N_i \ket i$
holds on a basis of states $\ket i$,
where $N_i$ is the sign of  $\bk{i}{i}$.
Then, our initial assumptions about $\hat{A}$ are equivalent to demand that $[\hat{A},\hat G_A]=0$ has one solution:
the unique ghost operator associated to $\hat{A}$ is  $\hat G_A\ket{A_i} = N_i \ket{A_i}$
 and allows to define a positive $A$-norm as $\bk{\psi'}{\psi}_A \equiv \bAk{\psi'}{\hat G_A}{\psi}$.  Then $A$-norm convergence agrees with coefficient convergence:
\beq
 \lim_{n\to \infty} \frac{ \bAk{\psi\np}{  (\hat P_i- p_i)^2}{\psi\np}_A   }{ \bk{\psi\np}{\psi\np}_A}  = 0,\qquad
p_i =  \frac{|\bk{i}{\psi}_A|^2}{\bk{\psi}{\psi}_A}
\eeq
In normal quantum mechanics all states have positive norm, and the ghost operator reduces to the unity operator.
With indefinite norm, the norm of a state contains one bit of information (its sign) which is preserved by time evolution
and affects a measurement trough the ghost operator.




\medskip

Does this prescription give an interpretation of \new{indefinite}-norm quantum mechanics that respects conservations laws?
Conserved quantities are associated to operators $\hat{A}$ that commute with the Hamiltonian $\hat H$.

We start discussing conservation of energy, associated with  $\hat H$ itself.
If dynamics is such that all eigenstates of $\hat H$ lie away from the `null cone',
then $[\hat H,\hat G_H]=0$, so that the extra factor $\hat G_H$ that appears in the interpretation of measurements of $\hat H$, 
does not spoil energy conservation.
As already mentioned, the same operator $\hat G_H$ is postulated to define probabilities in the context of
$PT$-symmetric quantum mechanics~\cite{Bender,Mosta}.

The same holds for any conserved observable $\hat{A}$: if $[\hat H,\hat{A}]=0$
one can find a common basis where $\hat{A}$ and $\hat H$ are simultaneously diagonal, 
such that the associated ghost parities coincide, $\hat G_A = \hat G_H$.
The sum of two commuting observables $\hat{A}$ and $\hat{A}'$ is interpreted additively.
In conclusion, the interpretation suggested by repeated states respects conservation laws.

\medskip

The situation becomes more interesting when interpreting observables $\hat{A}$ that are not conserved, $[\hat H,\hat{A}]\neq 0$.
Then the ghost operators associated to $\hat H$ and $\hat{A}$
\beq \hat G_H = \sum_i \kb{H_i}{H_i},\qquad 
\hat G_A = \sum_i \kb{A_i}{A_i}\eeq
are not equivalent under ${\rm U}(N_+, N_-)$ rotations that conserve the \new{indefinite} norm
(altought they would be equivalent under ${\rm U}(N_+ + N_-)$ rotations that conserve a positive norm,
where $N_+$ and $N_-$ is the number of positive-norm and negative-norm eigenstates).
This mismatch gives novel physical effects.
Time evolution is unitary in the \new{indefinite} norm; adding a factor $\hat G_A$ in the interpretation implies an extra non-conservation that only affects those observables which were already non conserved.
In section~\ref{2state} we discuss a simple example:
a number operator $\hat A$ in a `flavour' basis where the Hamiltonian is non diagonal.\footnote{In the language of
$PT$-symmetric Hamiltonians, this situation corresponds to setups where $C\neq P$.}

\subsection{Observables that don't commute with any ghost operator}\label{Anocom}




We next study observability of self-adjoint 
operators $\hat A$ that, like $\hat q$ and $\hat p$, posses some eigenvectors along the `null cone' 
of states with vanishing norm.
The identity 
\beq \langle A_i | \hat A |A_j  \rangle = \langle A_i | A_j\rangle A_j = A^*_i \langle A_i | A_j\rangle \eeq
implies that zero norm eigenvectors 
with  $\bk{A_i}{A_i}=0$ and $\bk{A_i}{A_j}\neq 0$
form pairs
with complex conjugated eigenvalues, $A_j = A_i^*$: 
in agreement with Bohr complementarity there is one real parameter per state.

For simplicity, let us consider a 2-state system.
The norm, written in terms of the two eigenvectors $\ket{0_\pm}$ of $\hat A$, is
\beq\label{eq:pmtheta}
\bk{0_{+}}{0_+} =\bk{0_{-}}{0_-} =0,\qquad
\bk{0_{+}}{0_-}= \bk{0_{-}}{0_+}  =1.
\eeq 
Both the norm and $\hat A=
A_{0_+}\kb{0_+}{0_-} + A_{0_-} \kb{0_-}{0_+}$ are  invariant under the transformation $\ket{0_+} = e^\theta \ket{0'_+}$, $\ket{0_-}= e^{-\theta^*}\ket{0'_-}$.
The complex parameter $\theta$ performs a U(1,1)  boost transformation, 
which acts diagonally on `null cone' states $\ket{0_\pm}$,
as can be verified by expressing them in terms of orthogonal states $\ket{\pm}$ with norm $\pm 1$
\beq \ket{0_+} = e^{ \theta} \frac{\ket{+} +  \ket{-}}{\sqrt{2}},
\qquad
\ket{0_-} = e^{- \theta^*} \frac{\ket{+} -  \ket{-}}{\sqrt{2}}.\eeq
This means that an operator $\hat A$ with eigenvectors along the null-cone 
commutes with U(1,1) rotations and thereby does not define a basis.
So the physical interpretation of a null-cone operator is analogous to the interpretation of the unit operator
in positive-norm quantum mechanics
(where the unit operator is the only operator that commutes with U(2) rotations
and that thereby does not define a basis):
$\hat A$ is a blind operator.
A generic state 
\beq\ket{\psi} = e^\lambda[ c_{0_+} \ket{0_+} + c_{0_-} \ket{0_-}]\eeq
can be rotated to any arbitrary vector (for example to $\ket{+}$ or to $\ket{-}$, depending on its norm) 
without affecting the observable associated to $\hat A$ by using the free projective
parameter $\lambda$ and the free boost parameter $\theta$.
Thereby there is no observable associated to $\hat  A$.

Indeed, the results of section~\ref{neversayprob} do not extend to operators with eigenvectors along the null-cone,
for the following reasons.
The freedom to rotate $\ket{\psi}$ is inherited by its repeated states $\kpn$, which fail to converge
towards a well defined state.
Furthermore, it is not possible to associate a ghost operator $\hat G_A$ to $\hat A$:
the equation $[\hat A,\hat G_A]=0$ has no solutions.
To verify this, let us try to define a ghost operator using the arbitrary $\ket{\pm}$ basis of eq.\eq{pmtheta}: 
\beq
\hat G_\theta = \kb{+}{+} \,+\, \kb{-}{-} = 
e^{2\Re\theta} \kb{0_-}{0_-} +e^{-2\Re \theta} \kb{0_+}{0_+}.\eeq
Both $\hat G_\theta$ and the associated positive norm $\bAk{\psi}{\hat G_\theta}{\psi} = e^{2\Re\theta} |c_{0_+}^2| + e^{-2\Re\theta} |c_{0_-}^2|$ depend on the arbitrary parameter $\theta$.
Furthermore, the tentative ghost operators $\hat G_\theta$ do not commute with $\hat A$
(unless $\hat A$ is the unit operator, namely if $A_{0_+} =A_{0_-}$ are real: this will be discussed in section~\ref{deg}).

\medskip

In the special case where $\hat A$ is the Hamiltonian, 
writing the pair of conjugated eigenvalues as $E_{0_\pm} = E \pm i \Gamma/2$
one can decompose $\hat H = E \hat 1 + \Gamma \hat L/2$ where $\hat L$ is the generator of U(1,1) boosts.
Thereby  $\hat U=e^{-i \hat H t}$ contains  $e^{\pm \Gamma t/2}$ factors that boost the kets, leaving physics unaffected.

\subsection{Observables that commute with many ghost operators}\label{deg}

Let us focus on the operator $\hat q$.
In the Pauli-Dirac representation 
\beq \label{eq:PD}
\hat q\ket{ x} =  i x \ket{ x},\qquad \bk{x'}{x}=\delta(x+x'),\qquad
 \bk{\psi'}{\psi} = \int_{-\infty}^{+\infty} dx \, \psi^{\prime *}(x) \psi(-x)\eeq
the eigenvectors  $\ket{x}$ of $\hat q$ have zero Pauli-Dirac inner product
and purely imaginary eigenvalues $ ix$, given that $x$ is real.
For each $|x|$ the eigenvectors form a  pair of zero norm states, $\ket{x}$ and $\ket{-x}$.
Their linear combinations
\beq \ket{\pm_x} = \frac{e^{-\theta(x)} \ket{x}\pm e^{-\theta(-x)} \ket{-x}}{\sqrt{2}}\eeq
have  diagonal inner product  $\bk{{\pm'}_{x'}}{\pm_x} = \pm\delta_{\pm'\pm} \delta(x-x')$
provided that
$\theta(-x) = -\theta^*(x)$.
Then the transformation
$\ket{x} \to  e^{-\theta(x)}\ket{x}$ 
is a local U(1,1) rotation at each value of $x$
which leaves the norm $\bk{x'}{x}$ invariant.
The $\ket{\pm_x}$ states are not eigenstates of $\hat q$,
so the ghost operators 
\beq \hat G_\theta = \int_0^\infty  dx  \, ( \kb{+_x}{+_x}+\kb{-_x}{-_x})\eeq
do not commute with $\hat  q$.
This is the situation discussed in the previous section. 

Nevertheless one can try to observe
pairs $\ket{\pm_x}$ of eigenvectors of $\hat q$, which is equivalent to observing $\hat q^2$.
The observable $\hat q^2$ is proportional to the 
unit operator within each pair, so that $\hat q^2$ commutes with all the $\hat G_\theta$ ghost operators.
The associated positive $\theta$-norms depend on the arbitrary function
$\theta(x)$: $\bAk{\psi}{\hat G_\theta}{\psi} = \int dx\, |\psi(x) e^{\theta(x)}|^2$, leading to an ambiguous interpretation.
The imaginary part of $\theta(x)$ 
corresponds to the usual freedom of locally re-phasing the states $\ket{x}$;
its real part provides extra freedom.


The ambiguity encoded in  $\theta(x)$ is eliminated by imposing that the $\hat p$ operator performs translations,
such that $\hat q$ eigenstates at different positions are related by
$\ket{x+\delta} = e^{-i\hat p \delta}\ket{x}$, which fixes $\theta(x)=0$,
such that $\hat G_0$ gives the positive $H$-norm.
The observable eigenvalue of the repeated $\hat q^{2(n)}$ operator acting over a repeated state $\kpn$ is then
$-\int dx\, x^2 |\psi(x)|^2/\int dx\,|\psi(x)|^2$.\footnote{The formalism invites to consider special theories where
$\theta(x)$ remains as a gauge redundancy
that combines local re-phasing and scale invariance.
This requires a complex extension of the vector potential.}

\section{Examples}\label{examples}
We now provide explicit examples of the discussion of section~\ref{negnorm}.

\subsection{The indefinite-norm two-state system}\label{2state}
We start from the simplest non-trivial quantum system, that consists of two energy eigenstates
$\ket{E_+}$ and $\ket{E_-}$ with norm $+1$ and $-1$,
such that time evolution is
\beq \ket{\psi(t)} = c_{E_+} e^{-iE_+ t} \ket{E_+} +c_{E_-} e^{-iE_- t} \ket{E_-} . \eeq
We assume that it is possible to measure a  `flavour' observable $\hat{A}$ with eigenstates
\beq\left\{\begin{array}{ll}
\ket{A_+} = \alpha \ket{E_+} + \beta^*\ket{E_-}   =\cosh \theta \ket{E_+} + \sinh\theta \ket{E_-}\cr
\ket{A_-} = \beta\ket{E_+} + \alpha^* \ket{E_-} = \sinh\theta \ket{E_+}+ \cosh \theta \ket{E_-} 
\end{array}\right.  .\eeq
The most generic U(1,1) rotation is parameterized by two complex numbers $\alpha,\beta$
subject to $|\alpha^2| - |\beta^2|=1$. Without loss of physical generality we can rephase $\ket{E_\pm}$
making $\alpha,\beta$ real, obtaining the second form in terms of one real boost parameter $\theta$.

\begin{figure}[t]
$$\includegraphics[width=0.45\textwidth]{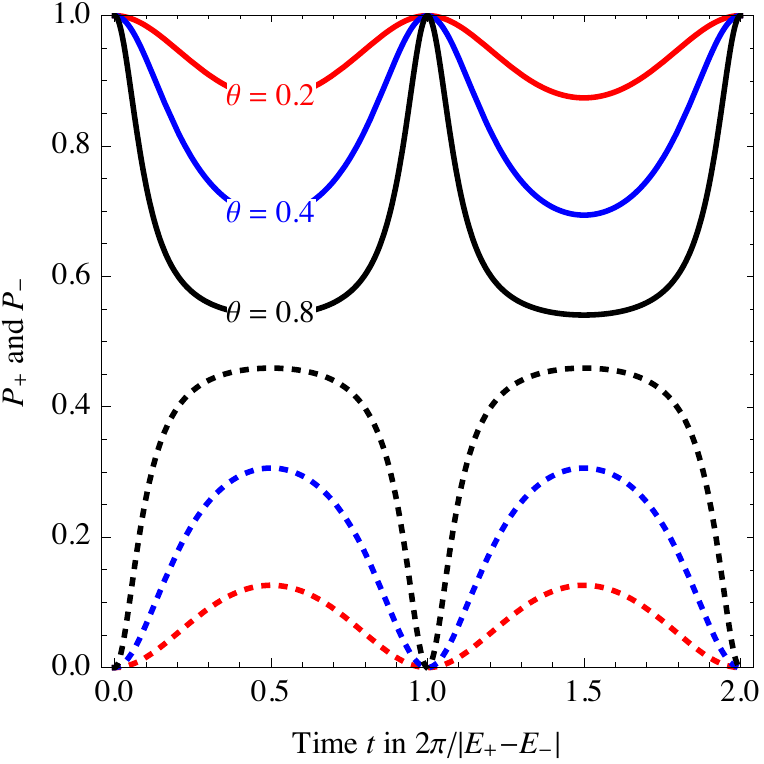},\qquad
\includegraphics[width=0.45\textwidth]{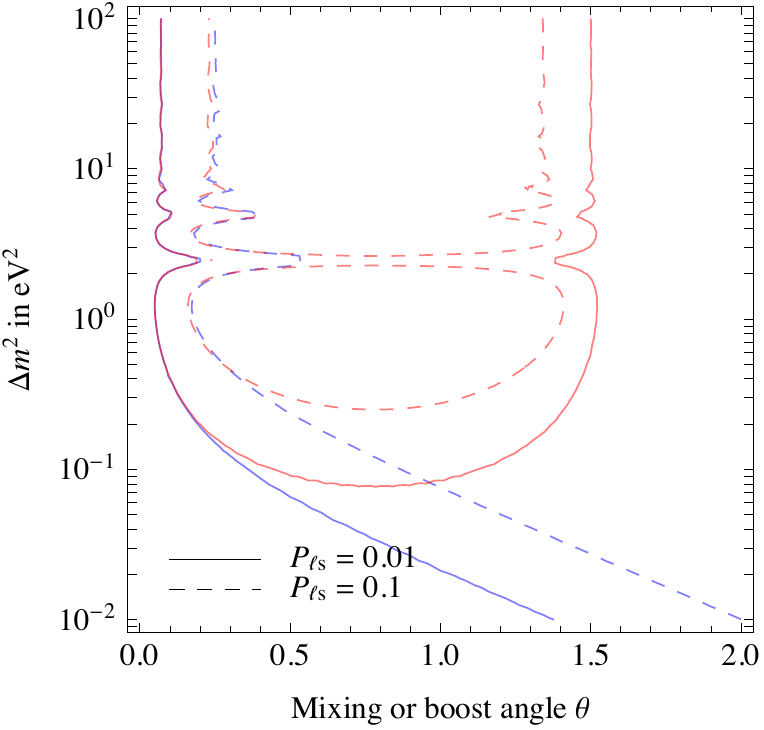}
$$
\caption{\label{fig:NegNormOsc}\em {\bf Left:} values
of $P_-(t)$ (continuous curves) and of $P_+(t)$  (dotted curves),
starting from the state $\ket{A_-}$ at $t=0$.
{\bf Right:} contours with active-to-sterile neutrino oscillation probability
equal to 0.01 (continuous), 0.1 (dashed) for normal oscillations (red) and for oscillations into negative-norm
sterile neutrino (blue). We assumed $L/E = {\rm km}/\GeV$.
}
\end{figure}

Assuming the initial state $\ket{\psi(0)}=\ket{A_-}$, we
compute the `oscillation' rate to the state $\ket{A_\pm}$ at time $t$.
Standard manipulations give\footnote{Similar equations are found in
studies of $PT$-symmetric hamiltonians~\cite{Bender,quant-ph/0609032}; however 
their physical goal and meaning is not clear, or at least different for different authors.
Some authors view $PT$-Hamiltonians as ordinary quantum mechanics written in a basis
that seems to give different effects, and interpret non-standard results such as~\cite{1312.3395} as the effects of some `apparatus' that switches on some interaction that changes the basis of eigenstates.
Given that all SM particles obey standard quantum mechanics, we don't know how this is possible.

Our motivation is the possibility that new physical particles with negative norm might exist.
In particular, 4-derivative gravity predicts a massive spin-2 ghost,
which might make sense if quantised with positive energy and \new{indefinite} norm.
Another possible physical application is neutrino oscillations into a  speculative new sterile state with negative norm. 
}
\beq P_{+} = \frac{\sin^2(\frac12 (E_+-E_-) t) }{\coth^2 2\theta - \cos((E_+-E_-)t)}
\label{eq:PNN}
,\qquad
 P_{+} + P_- = 1 .\eeq
The non-standard shape of the \new{indefinite}-norm oscillation rates $P_\pm(t)$ is plotted in fig.~\ref{fig:NegNormOsc};
its time average is $\langle P_+\rangle = 1/[2 (1- \cosh^{-1/2}4\theta)]$.
The bound $ P_- \ge 1/2$  holds also in the case where one negative-norm state
interacts with an arbitrary number of positive-norm states, given that the \new{indefinite} norm is conserved by time evolution,
with implications for stability of the lightest negative-norm particle.

Considering neutrino oscillations into a  speculative new sterile state with negative norm,
the oscillation probabilities of eq.\eq{PNN}
have the same physical meaning as the usual neutrino oscillation probabilities.
A qualitatively new feature is oscillations with $\theta \gg 1$,
such that sizeable transition probabilities can arise even at small values of the oscillation phase, see
fig.\fig{NegNormOsc}b.

\medskip

As a possible physical application, we recall that the `$3+1$' scheme (3 active neutrinos plus a sterile neutrino, all with positive norm)
cannot  fit the LSND and {\sc MiniBoone} $\nu_\mu\to \nu_e$ anomaly~\cite{1805.12028} because
$\nu_e$ and $\nu_\mu$ disappearance experiments imply too strong bounds~\cite{3+1}.
It is interesting to check the viability of a   `$3-1$' scheme (3 active neutrinos plus a negative-norm sterile neutrino),
given  the difference in the oscillation formula.
Writing the most splitted neutrino mass eigenstate as
\beq \nu_1 = \nu_s\sqrt{1+\sum_\ell \sinh^2\theta_{\ell s}} + \sum_\ell  \nu_\ell \sinh\theta_{\ell s},\qquad
\ell=\{e,\mu,\tau\} \eeq
all other elements of the neutrino mixing matrix follow from unitarity, and
the relevant oscillation probabilities are
\beq
P_{ee}=\frac{1+S\sinh^2 2\theta_{es}}{1-4S \sinh^2\theta_{es}(1-\sum_\ell \cosh2\theta_{\ell s})},\qquad
P_{\mu e}=\frac{4S\sinh^2\theta_{e s}\sinh^2\theta_{\mu s} }{1-4S \sinh^2\theta_{\mu s}(1-\sum_\ell \cosh2\theta_{\ell s})}\eeq
where $S = \sin^2 (\Delta m^2 L/4E_\nu)$ is the usual oscillation factor.
$P_{\mu\mu}$ and $P_{\mu e}$ are obtained by permutations; notice that $P_{\mu e}\neq P_{e\mu}$.
For small `angles' $\theta_{\ell s}\ll 1$ the oscillation probability reduces to the one of 3+1 oscillations.
We verified that the $3-1$ scheme has problems analogous to the $3+1$ scheme in fitting the anomaly.

\subsection{The indefinite-norm free harmonic oscillator}\label{2der}
A 4-derivative oscillator can be decomposed as two modes: one with positive classical Hamiltonian, and one
`ghost' with negative classical Hamiltonian $\hat H=-\frac12 (\hat q^2+\hat p^2)$ (see e.g.~\cite{1512.01237}).
As a second example, we focus on the `ghost' and quantise it
using the Dirac-Pauli representation for $\hat q$ and $\hat p$ discussed in the introduction.
The combination of the unusual factors leads to the same Schroedinger equation as for the positive harmonic oscillator.
The energy eigenvalues are as usual $E_k = k+1/2 \ge 0 $ for integer $k\ge 0$.
The eigenstates $\ket{E_k}$ have the usual bounded wave-functions.
The difference is that $q,p$ are self-adjoint with respect to the indefinite norm  of eq.\eq{PD},
which is negative on anti-symmetric wave-functions, leading to $\bk{E_{k'}}{E_k}=(-1)^k \delta_{k k'}$.

\smallskip

The same result can be re-obtained 
rewriting the Hamiltonian as $\hat H= -\frac12 (\hat a \hat a^\dagger  + \hat  a^\dagger \hat a)$ where
$\hat a= (\hat q+i\hat p)/\sqrt{2}$, $\hat a^\dagger = (\hat q-i \hat p)/\sqrt{2}$ and $\dagger$ denotes self-adjoint
with respect to the indefinite norm implied by 
$[\hat a, \hat a^\dagger]=-1$ and $\hat a\ket{0}=0$.\footnote{\new{In our notation
the  standard positive-norm quantization would correspond to the alternative vacuum $\hat a^\dagger\ket{0_{\rm st}}=0$
(usually described, in standard notation, by swapping $\hat a$ and $\hat a^\dagger$).  
The classical limit of the positive-norm quantum theory~\cite{Osc+,astro-ph/0601672} respects the correspondence principle, being a
potentially problematic negative-energy theory~\cite{Ostro,AntiOstro}}.}
This gives $\ket{E_k}=(\hat a^\dagger)^k\ket{0}/\sqrt{k!}$ as eigenstate of $\hat H$ with positive eigenvalue 
$E_k = k+1/2$ and norm $(-1)^k$.
\new{Wave-functions are normalizable, solving one issue raised in~\cite{astro-ph/0601672}}.
This exemplifies how dynamics determines an indefinite norm.

The ghost operator $\hat G_H$ associated to $\hat H$ is parity,\footnote{In Quantum Field Theory
  it becomes parity in field space.}
which flips $\hat q, \hat  p$,  so that $\{\hat G_H , \hat q\}=0$.
In view of the half-integer values of $E_k$, wave functions flip sign in one period,
and the $H$ ghost operator can be written as $\hat G_H = i e^{-i \pi \hat H} = i \hat U(\pi)$ where $\hat U(t)$ is
the evolution operator, and $\pi$ corresponds to a half-period. 

\begin{figure}[t]
$$\includegraphics[width=0.46\textwidth]{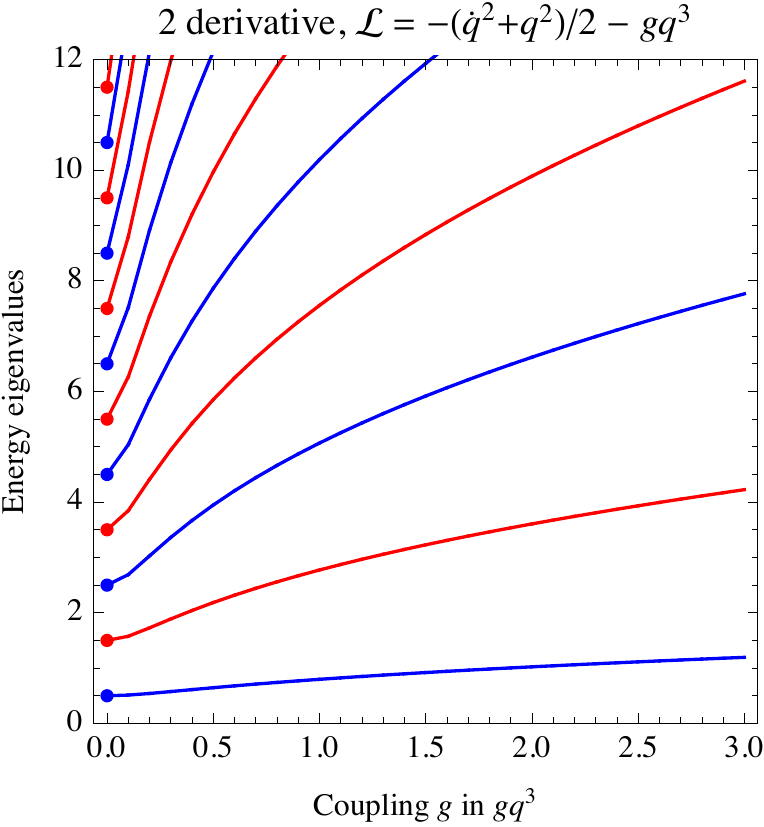}\qquad
\includegraphics[width=0.46\textwidth]{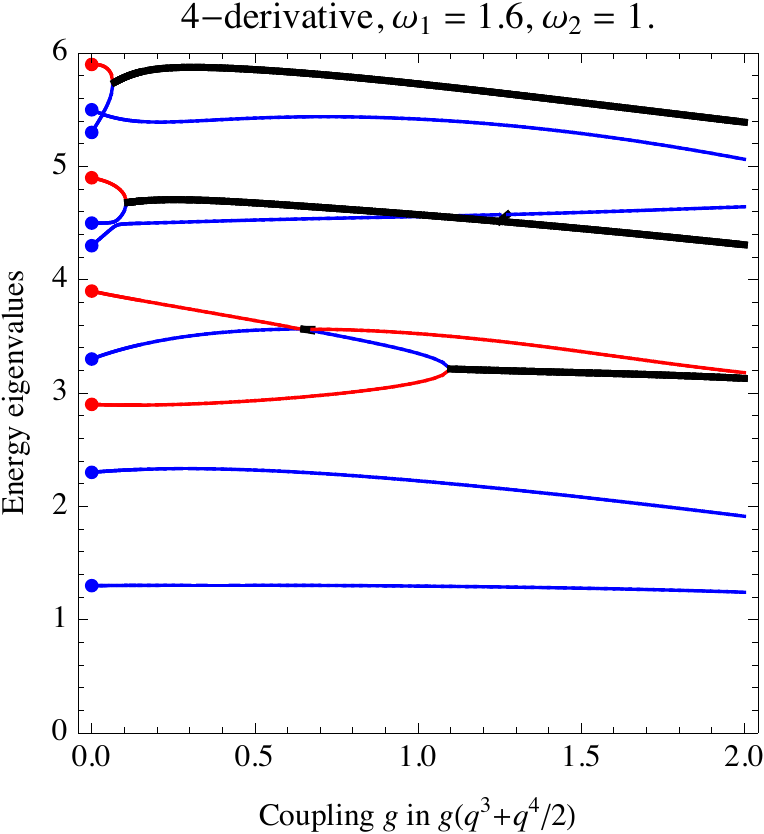}$$
\caption{\label{fig:Ei}\em Values of the energy eigenstates with positive norm (blue),
negative  norm (red) and pairs of zero norm (black thick) 
for a 2-derivative oscillator
in the presence of an interaction $gq^3$ (left) 
and for a 4-derivative oscillator in the presence of an interaction $g(q^3+q^4/2)$ (right).}
\end{figure}

\medskip

We next add interactions,  e.g.\ a potential 
$V(q) =kq - \frac12 q^2 + g q^3 + \lambda q^4$.
Interactions that respect parity (even powers of $\hat q$) 
do not affect the ghost operator and can be treated along the lines of~\cite{1611.03498}.\footnote{In Quantum Field Theory
this correspond to negative-norm particles that only couple in pairs.
This assumption does not apply to 4-derivative gravity (see section~\ref{4der}).}

If instead $V$ breaks parity, the ghost operator $\hat G_H$ is no longer parity. 
The special case of a linear term $kq$ is analytically solved by a shift in $q$, 
such that $\hat G_H$ becomes shifted parity.
In fig.~\ref{fig:Ei}a we show the energy eigenvalues obtained adding a $gq^3$ interaction.
For small $g$ the eigenvalues can be computed perturbatively starting from the basis of the free oscillator.
For large $g$ we use the Pauli-Dirac coordinate representation discretised on a space lattice.
Energy eigenstates remain away from the `null cone':
like for the harmonic oscillator, all eigenstates of $\hat H$ have either positive or negative norm
and there are no null kets.


\subsection{The 4-derivative oscillator}\label{4der}
We next consider an interacting 4-derivative variable $q(t)$
with Lagrangian 
\beq {\cal L} = - \frac12 q (\frac{d^2}{dt^2} +\omega_+^2)(\frac{d^2}{dt^2} +\omega_-^2)q - g q^3 -\lambda q^4.\eeq
It can be decomposed into two interacting 2-derivative modes modes $q_1 = q$ and $q_2 = \dot q$.
In the free limit ($g=\lambda=0$) it is more convenient to use 
the alternative canonical variables $ q_{\pm}$ defined by
\beq q= \frac{{q}_+   + {q}_-}{\sqrt{\omega_-^2 - \omega_+^2}},\qquad
\ddot q = -\frac{\omega_-^2  q_- + \omega_+^2  q_+}{\sqrt{\omega_-^2 - \omega_+^2}}.
\eeq
Indeed, $ q_\pm$ are two $2$-derivative decoupled harmonic oscillators with
frequencies $\omega_{\pm}$ and signs $\pm$ of their classical energies.
In the presence of the couplings, the two modes $ q_\pm$ interact such that 
classical solutions exhibit sick run-away behaviours.\footnote{Some higher derivative theories
avoid run-away solutions, for a range of initial conditions~\cite{1607.06589}.}

\smallskip


We now study the quantum theory.
As described in the introduction, time-reflection demands using
the Schroedinger coordinate representation for $\hat q_1$ and
the Pauli-Dirac coordinate representation for $\hat q_2$.
We discretise the Hamiltonian on a lattice in $(q_1, q_2)$ and compute its eigenvalues and eigenvectors.
The  energy eigenvalues are shown in fig.~\ref{fig:Ei}b as function of $g$ assuming
 the indicated values of $\omega_{\pm}$ and $\lambda = g/2$.
The corresponding wave-functions $\psi(q_1,q_2)$ of $\hat H$ are normalizable.
In the presence of interactions, some energy eigenstates form zero-norm states.

\medskip

As discussed in the introduction, a 4-derivative variable $q(t)$ 
is a useful toy model for a 4-derivative graviton field $g_{\mu\nu}(x)$:
$ q_+(t)$ mimics massless graviton modes with frequency $\omega_+ = p$,
 where $p$ is the spatial momentum;
$ q_-(t)$ mimics massive graviton ghost modes with $\omega_- = \sqrt{p^2 +M^2}$.
The graviton has a linear coupling $g_{\mu\nu} T_{\mu\nu}$ to the energy-momentum tensor of matter fields.
We mimic it adding a coupling $q j$, where $j\sim f^2$ is some function of  `matter variables' $f(t)$.

When computing virtual $q$ exchange between matter states, 
the sum over $q\propto {q}_{+}+q_-$ reproduces the 4-derivative propagator
\beq \label{eq:qprop}
\frac{1}{\omega_-^2-\omega_+^2} \bigg[\frac{1}{\omega^2-\omega_+^2} - \frac{1}{\omega^2 -\omega_-^2}\bigg] = 
-\frac{1}{ (\omega^2-\omega_-^2)(\omega^2-\omega_+^2) }. \eeq
This result can also be obtained from 2nd order perturbation theory applied to a toy 3-state system:
$\ket{j}$, that couples to $\ket{+} + \ket{-}$, with free energies $\omega, \omega_+$ and $\omega_-$ respectively.
The cancellation in eq.\eq{qprop} implies renormalizable  gravitational $ff\to ff$ cross sections at large energy
$\omega \gg M$.
The fact that external scattering states have positive norm bypasses the issue of interpreting negative-norm states~\cite{LeeWick}.

We can now address this issue. 
Matter couples in the same way to the two graviton components,
so the observable that discriminates $q_+$ from $q_-$ is their different invariant mass.
Energy eigenstates are affected by interactions with matter states.
Their effects can be encoded in an effective Hamiltonian $H_{\rm eff} = \diag(\omega_+,\omega_-) + \delta H$
restricted to the $q_\pm$ system.
In Wigner-Weisskopf approximation $\delta H$ takes the time-independent  form
$\delta H = \delta M - i \Gamma/2$, where $\delta M$ and $\Gamma$ are self-adjoint.
Since matter couples to $q_+  + q_-$,
all matrix elements of $\delta H$ are equal at large energies $E \gg M$,
where we can neglect the mass difference between $q_+$ and $q_-$.
The $\delta H$ term alone is the critical Hamiltonian with degenerate
eigenvectors that appears when eigenstates with norm $\pm 1$ become degenerate,
forming a null-norm pair~\cite{1512.01237}.
Diagonalising the full $H_{\rm eff}$ at different virtual energies $E$
gives graviton eigenstates
$\ket{q_\pm (E) } = \cosh\theta(E)\ket{q_\pm} + \sinh\theta(E) \ket{q_\mp}$ 
with norm $\pm$ that,
at energies above $M$,  asymptotically rotate towards the null-cone, $\theta \to \infty$ as $E\to \infty$.
In our interpretation, this rotation implies a suppression $e^{-\theta} \sim [\omega_\pm/\delta H]^{1/4}$ of graviton couplings to matter.

Measurements on  states that entangle positive with negative norm 
(for example produced in $ff\to {q}_\pm {q}_{\pm'}$ or $f'\to q_\pm f$ processes)
seem to allow to transmit information at space-like distances~\cite{1312.3395},
and for related phenomena~\cite{quant-ph/0609032}.
This is not possible in ordinary quantum mechanics, where this feature follows
from locality and relativity~\cite{quant-ph/0212023}.
This issue will thereby be better addressed in
relativistic quantum field theories with \new{indefinite} norms:
is it possible to define covariant ghost operators that commute with field observables at space-like distances?

\section{Conclusions}\label{concl}
In section~\ref{neversayprob} we discussed how the deterministic part of the Born postulate
({\em ``when an observable corresponding to a self-adjoint operator $\hat A$ is measured in an eigenstate $\ket{A_i}$,
the result is the eigenvalue $A_i$''}) allows to deduce the full Born postulate,
that gives a probabilistic interpretation of generic states $\ket{\psi}$.
As illustrated in fig.~\ref{fig:bells},
repeated states
$\kpn = \ket{\psi}\otimes \ket{\psi} \otimes \cdots  \ket{\psi}$ 
become, for large $n$, eigenstates of $\hat A\np$: the observable that measures the average of $\hat A$
over $n$ repeated experiments.
The deterministic part of the Born postulate then predicts the outcome of the average,
reproducing the usual result in terms of frequentist probability.
The limit $n\to\infty$ is defined in two different ways
(norm convergence and coefficient convergence),
which give the same result.
The  rigid mathematics of quantum theory forces its interpretation.

\medskip

Dirac and Pauli have shown that a pair of self-adjoint canonical operators
$\hat q$ and $\hat p$ admits, beyond the usual Schroedinger coordinate representation,
a different representation that leads to an indefinite norm.
In section~\ref{negnorm} we used repeated
states and measurements to search for an interpretation of quantum mechanics with indefinite quantum norm.
The two different limiting criteria that can decide whether $\kpn$ becomes an eigenvector of $\hat A\np$  in the limit $n\to\infty$
are no longer equivalent.
The first criterium, norm convergence, does not lead to a probabilistic interpretation and does not imply coefficient convergence,
given that the norm is indefinite.
The second criterium, coefficient convergence, implies a successful probabilistic interpretation for 
a class of self-adjoint operators $A$.

The intuitive result following from repeated states and measurements is formalised 
in a way similar to $PT$-symmetric Hamiltonians~\cite{Bender,quant-ph/0609032,Mosta,1611.03498}.
We define a `ghost operator' $\hat G$ as a linear operator that acts as $\hat G\ket{i} = N_i \ket{i}$
on a basis of states $\ket{i}$, where $N_i$ is the sign of $\bk{i}{i}$.
This allows to define an artificial positive norm as $\bAk{\psi'}{\hat G}{\psi}$;
however both $\hat G$ and its associated positive norm are basis-dependent.

We try to associate to any observable $\hat A$ a  ghost operator $\hat G_A$ such that \beq\label{eq:appAG} [\hat A,\hat G_A]=0.\eeq
Three different situations are encountered:
\begin{itemize}
\item[1)] If eq.\eq{appAG} has a unique solution, $\hat G_A$ allows to define a positive norm 
and a probabilistic interpretation of $\hat A$, which agrees with the interpretation suggested by repeated measurements.
As discussed in section~\ref{neversayprob}, the solution is unique when the eigenvectors of $\hat A$ define a unique basis:
the eigenvectors $\ket{A_i}$
must lie away from the null-cone of configuration space, and the eigenvalues $A_i$ must be non-degenerate.
\end{itemize}
The ghost operator contains one  bit of information preserved by time evolution
(the sign of the norm of $\bk{\psi}{\psi}$) and
reduces to the unity operator when the norm is positive.
In general,  the ghost operator is observable-dependent, giving rise to unusual effects
for observables that do not commute with $\hat H$, such as an unusual oscillation probability
of positive-norm states into negative-norm states, as discussed in section~\ref{2state} and fig.~\ref{fig:NegNormOsc}.
One possible physical application is neutrino oscillations into a speculative sterile neutrino with negative norm.

\begin{itemize}
\item[0)] If no ghost operator commutes with $\hat A$, we cannot associate any positive norm to $\hat A$ and no interpretation.
In section~\ref{Anocom} we show that 
this happens for operators with pairs of eigenvectors along the null-cone of configuration space.
In such a case $\hat A$ commutes with  norm-preserving U(1,1) `boosts' 
in the 2-dimensional subspace, given that `boosts'   act multiplicatively along the null-cone.
Thereby such operators fail to define a basis in configuration space.
In more physical terms, they correspond to measurement apparata that cannot split states into events.
This is the case of the $\hat q$ position operator in Pauli-Dirac coordinate representation.

\item[2)]
If eq.\eq{appAG}  has multiple solutions, the interpretation of $\hat A$ is ambiguous. 
This is the case of degenerate
operators, which have the same eigenvalue for a positive-norm state $\ket{+}$ and a negative-norm state $\ket{-}$.
This describes measurements that do not discriminate $\ket{+}$ from $\ket{-}$,
as they manifest as the same macroscopic state.
As discussed in section~\ref{deg}, the Dirac-Pauli  $\hat q^2$ operator belongs to this category:
it does not discriminate $\ket{x}$ from $\ket{-x}$.
In such a case the ambiguity is removed  by imposing that $\hat p$ acts as translations,
dictating the interpretation of $\hat q^2$.

\end{itemize}
The relation between $\hat q$ and $\hat q^2$ is reminiscent of fermions fields $\Psi$, which are not observable unlike $\bar\Psi \Psi$ bilinears.
In particular, the free harmonic oscillator with Dirac-Pauli coordinate representation can be exactly solved:
eigenstates have norm and parity $(-1)^k$, such that the ghost operator $\hat G_H$ associated with the Hamiltonian is parity: $\hat q$ and $\hat p$ anti-commute with $\hat G_H$.
The eigenvalues are $E_k =k+1/2$ with integer $k\ge 0$, such that
wave-functions flip sign in a period.
The classical limit of a \new{indefinite}-norm harmonic oscillator can be explored trough coherent states,
but we have not extracted a general lesson from this simple system.
Interactions are considered numerically in section~\ref{2der}.
\mio{The classical limit of a harmonic oscillator is [somehow]
an oscillator shifted by one quarter of period,
but we do not know a general answer.}

\medskip

In section~\ref{4der} we studied an interacting 4-derivative variable $q(t)$.
This toy model will allow to address 
dimension-less renormalizable theories of quantum gravity where 4-derivatives act on the graviton,
which give an extra heavy graviton as the only negative-norm particle.

\subsubsection*{Acknowledgments}
This work was supported by the ERC grant NEO-NAT.
A.S.\ thanks Nima Arkani-Hamed, Kenichi Konishi, Alberto Salvio and Riccardo Torre for useful discussions.

\footnotesize


  \end{document}